\documentstyle[11pt,epsf,paspconf]{article}

\def\gi{{\it Ginga}}
\def\ein{{\it Einstein Observatory}}

\def\ros{{\it ROSAT}}
\def\xte{{\it RXTE}}
\def\asca{{\it ASCA}}

\def\aox{{$\alpha_{\rm ox}$}}

\def\etal{{et al.}}

\def\lx{{$L_{\rm x}$}}
\def\lo{{$L_{\rm o}$}}

\def\fek{{Fe K$\alpha$}}

\begin{document}

\title{X-ray dependencies on luminosity in AGN}
\author{K. Nandra}
\affil{LHEA, NASA/GSFC, Greenbelt, MD 20771, USA}

\begin{abstract}
Several X-ray properties of active galactic nuclei depend, or appear
to depend, on their luminosity. It has long been suggested that \aox,
the X-ray ``loudness'' decreases with luminosity. There never has been
a satisfactory explanation of this observational claim, and the
statistical soundness of the result has been disputed. The earliest
systematic studies of the X-ray variability of AGN showed that these
properties also depend on luminosity. In particular, the normalization
of the power spectrum, or alternatively the the variability amplitude,
are anti-correlated with luminosity.  Most recently, tentative
evidence from \gi\ for an X-ray Baldwin effect - a decrease in the
\fek\ equivalent width with luminosity - has been confirmed and
extended by \asca. The new data show that the reduction in strength is
accompanied by changes in profile. These results will be described and
their interpretation discussed.
\end{abstract}

\keywords{X-rays, variability, black holes, accretion disks}

\section{Introduction}

Active Galactic Nuclei (AGN) are strong X-ray emitters.  The apparent
power-law nature of the X-ray continuum suggests a non-thermal
radiation mechanism. This demands an energy-deposition process which
is distinct and in addition to the thermal emission usually attributed
to an accretion disk (e.g., Malkan \& Sargent 1982). The X-ray
continuum of AGN is rapidly variable, suggesting a very compact region
close to the central black hole. Recent X-ray spectral observations
have confirmed this, by detecting the relativistic signatures of the
hole in the profile of the \fek\ line (Tanaka \etal\ 1995; Nandra
\etal\ 1997). Gravitational and Doppler effects distort the profile of
the line in a characteristic manner (Fabian \etal\ 1989; Laor 1991).
The X-ray emission of AGN does not therefore merely invite
explanation, but also offers an opportunity to examine the central
engine in detail.

An increasing number of the X-ray properties of AGN have been found to
be related to other parameters. For example, a menagerie of correlated
optical properties (Boroson \& Green 1992) appear also to be related
to the X-ray emission (e.g. Boller, Brandt \& Fink 1996; Lawrence
\etal\ 1997; Laor \etal\ 1997). The origin of these correlations is
currently the subject of intense study and speculation.  Also, certain
X-ray properties of AGN are correlated with luminosity. These
relationships are likely to be of great relevance to the physics of
AGN. Here, three X-ray properties of AGN which are, or appear to be,
related to source luminosity will be discussed. Evidence exists that
the following quantities decrease with luminosity:

\begin{itemize}
\item
The X-ray ``loudness''
\item 
The amplitude of X-ray variability on a fixed time scale
\item
The equivalent width of the \fek\ line
\end{itemize}

In the following sections, the evidence for and nature of these
correlations will be reviewed and speculation made on their origin and
sources of scatter. The prospects for improvement in the observational
data and theoretical interpretations will also be explored. Thereafter, 
the correlations will be examined for suitability as cosmological 
calibrators. 

\section{The Correlations}

\subsection{X--ray ``loudness'', \aox}

Data from the \ein\ gave the first indication that the X-ray emission
of AGN was correlated with luminosity. Early observations showed that
the ratio of the X-ray to the optical luminosity, \lx/\lo\ of quasars
depended on \lo (Avni \& Tananbaum 1982; Kriss
\& Canizares 1985). A convincing demonstration of this effect was
given by Avni \& Tananbaum (1986) who studied the PG sample of quasars
and a number of other, more heterogeneously-defined samples. These
workers showed that the spectral slope between 2500\AA\ and 2 keV,
\aox, increased with optical luminosity. This
steepening of the slope implies that the X--ray emission gets
relatively weaker as the optical luminosity increases. Their best
estimate for the relationship between optical and X-ray luminosity was
$L_{\rm x} \propto L^{0.8}_{\rm o}$. This correlation has also been
observed in a larger IPC sample (Wilkes \etal\ 1994) and in a large
\ros\ sample (Green \etal\ 1995), who found that \aox$\propto
L^{-0.1}_{\rm o}$.

%\begin{figure}
%\plotone{fig_aox.ps}
%\caption{Dependence of the optical-to X-ray spectral slope
%($\alpha_{\rm ox}$) on optical luminosity, presented by
%Avni \& Tananbaum (1986). The X-rays appear to get progressively
%weaker relative to the optical, as the source luminosity
%increases.
%\label{fig:aox}}
%\end{figure}

\subsubsection{The origin of the correlation}

The question of why the X-ray loudness might depend on the optical
luminosity has never really been answered.  Crudely, in the standard
model, one might expect the $L_{\rm o}$ to indicate the luminosity of
the accretion disk. This would depend on the accretion rate, the mass
of the black hole and the efficiency by which the disk converts the
rest-mass energy of the accreting material into radiation.  The
broad-band X-ray luminosity, on the other hand, should also depend on
the first two parameters but the relevant efficiency in this case is
that of energy deposition into the X-ray producing particle
distribution. \aox\ may therefore be revealing something about the
ratio of these efficiencies. There are, however, a number of other
considerations which may confuse this naive interpretation.

First of all, most AGN samples exhibit a strong correlation of
luminosity with redshift. Simple bandpass effects might therefore
induce artificial correlations. This seems unlikely in the present
case, as no correlation has generally been found between $\alpha_{\rm
ox}$ and red shift. There are also likely to be effects due to the
disk inclination. The apparent luminosity of the disk depends on its
orientation to the line-of-sight and if the X-ray luminosity does not,
or has a different dependence, this could be responsible for the
correlation. Another possibility is that X-ray absorption is more
common in AGN with high optical luminosity, espcially relevant as the
\ein\ IPC and \ros\ PSPC data are in the soft X-ray band, This 
is not thought to be the case, indeed the opposite is anecdotally
assumed, but no systematic studies have yet been
undertaken. Conversely, optical redenning could be more common in low
luminosity AGN.
An even more depressing interpretation of the observed correlation is
that it is due to a statistical effect. Yuan, Siebert \& Brinkmann
(1998) have used Monte-Carlo simulations to show that if the
distribution of \lo\ has a higher dispersion than that of \lx, an
apparent anticorrelation between \aox\ and \lo\ can be observed 
without any physical reason.

\subsubsection{Sources of scatter}

There is a lot of scatter in the \lx--\aox\ relation. While it was
mentioned above that absorption might be the cause of the observed
effect, if it were not, it could certainly be a cause of scatter.
Hard X-ray selected samples, which should be less affected by
absorption, show a wide range of absorption columns, covering several
orders of magnitude in column density (Turner \& Pounds 1989; Nandra
\& Pounds 1994). Such columns would have a dramatic effect on the soft
X-ray flux and therefore the inferred luminosity.

\begin{figure}[t]
\plotone{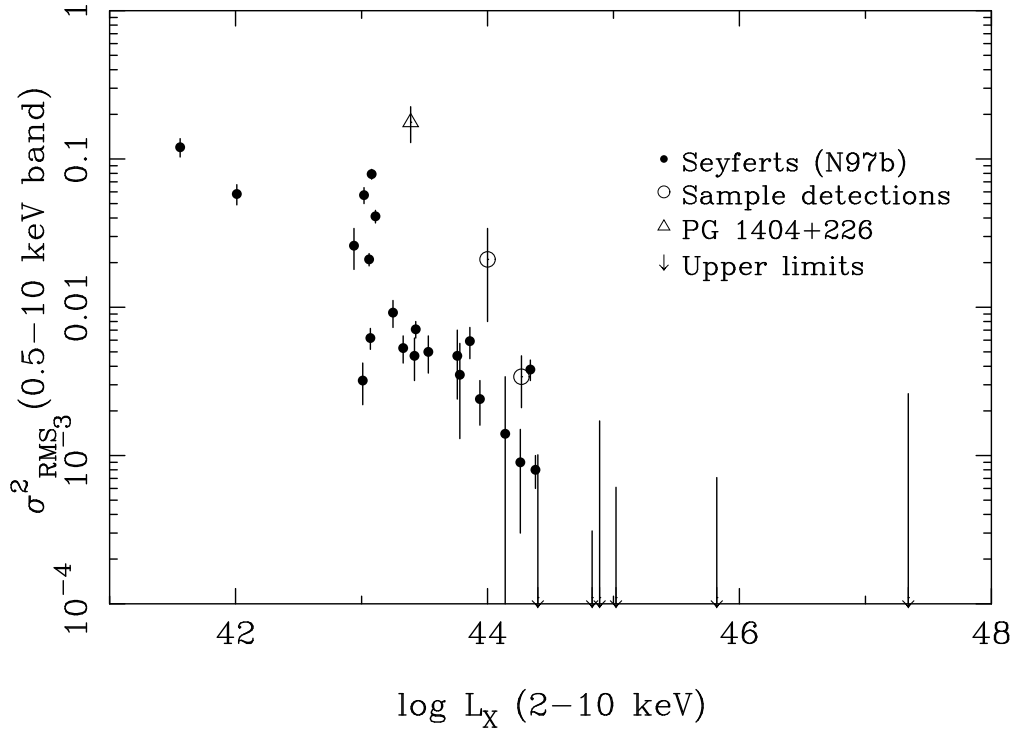}
\caption{
Dependence of the normalized ``excess variance'' $\sigma^{2}_{\rm
RMS}$, which measures the X-ray variability amplitude, versus X-ray
luminosity. The data are for a sample of Seyfert galaxies and
quasars observed by \asca\ (Nandra \etal\ 1997a). There is a
very strong anticorrelation, with an index of $\sim -0.7$. These
data confirm the previous {\it EXOSAT} results
\label{fig:rms}}
\end{figure}

\subsubsection{Future prospects}

The results on \aox\ have been established primarily with
optically-selected AGN, using soft X--ray data.  For the future, it
would be interesting to see if the correlations held for the hard
X-ray band, which is much less affected by absorption. It is also
important to consider samples which have been selected in different
ways. Further examination of both kinds of correlation may well be
possible with \asca, if large, well-selected samples are completed.

\subsection{X-ray Variability parameters}

It has long been suspected that higher-luminosity sources are, in some
sense, ``less variable'' in the X-ray band. For example, Barr \&
Mushotzky (1986) correlated a measure of the source doubling time
scale with luminosity and found a significant correlation. More
convincing evidence for such a correlation came from {\it EXOSAT}
observations (Lawrence and Papadakis, 1993; Green, McHardy \& Lehto,
1993). The power density spectra (PDS) of the {\it EXOSAT} AGN was
were found to be consistent with a single form, but the amplitude
showed a strong anticorrelation with the X-ray luminosity.

This result has been confirmed using \asca\ data by Nandra \etal\
(1997a). The normalized ``excess variance'', $sigma^{2}_{\rm RMS}$, of
an \asca\ sample of AGN is plotted against\lx\ in
Fig.~\ref{fig:rms}. $sigma^{2}_{\rm RMS}$ is defined as the variance
above that expected from the (Poissonian) errors in the data,
normalized by the mean count rate. This quantity should, for a
stationary PDS, represent the integral over the frequencies
corresponding approximately to the length of the observation and the
bin size. For a red-noise PDS, as observed in AGN, $\sigma^{2}_{RMS}$
depends on and increases with the observation length. The length of
the \asca\ observations was similar, however, and making accurate PDS
estimates for data which is unevenly-sampled is notoriously
difficult. It is therefore preferable to use this quantity, rather
than attempt to estimate the PDS amplitude.  The $\sigma^{2}_{\rm
RMS}$ anticorrelation found with \asca\ depends on $L_{\rm X}^{-0.71}$
(Table~\ref{tab:summary}). Similar dependencies $L_{\rm X}^{-0.68}$
and $L_{\rm X}^{0.55}$ were reported by Green \etal\ and Lawrence \&
Papadakis. The dependence is therefore much stronger than that of,
e.g., \aox, or the UV Baldwin relations (Baldwin 1977).

\subsubsection{The origin of the correlation}

Lawrence \& Papadakis (1993) briefly discuss some possible origins of
the correlation. The most obvious conclusion is that it is related to
the source size, although one might then expect a $L_{\rm X}^{-1}$
dependence. Bao \& Abramowicz (1996) have suggested a model in which
the X-ray variations are produced by hot-spots on a rotating accretion
disc. In that model, \lx\ depends on the inclination: more edge-on
disks have lower luminosities due to projection effects. The
variability is also enhanced due to increased relativistic
effects. This models conflicts somewhat with observations of the \fek\
line, which suggest that in type 1 Seyfert galaxies the accretion
disks are all observed close to face-on (Nandra \etal\ 1997b).  It is,
however, the only serious model which has been suggested so far to
explain the correlation.

\subsubsection{Sources of scatter}

Once again, it is clear that the correlation between the variability
parameters and the luminosity is imperfect. One possibility is that
the variability amplitude is also related to the so-called
``eigenvector 1'' of Boroson \& Green (1992). Their work revealed a
number of correlations between various optical properties of AGN and
some X-ray properties are also related. For example, narrow H$\beta$
emission, strong Fe {\sc ii} lines, steep soft X-ray spectra and large
amplitude, rapid X-ray variability all seem to be found in the same
objects, the so-called ``narrow-line Seyfert 1 galaxies'' (NLS1). A
systematic study of NLS1 by \asca\ (K. Leighly, priv. comm.) has
revealed that, as a class, these sources also follow the
$\sigma^{2}_{\rm RMS}$ vs. $L_{\rm X}$ relation, but with a higher
overall variability amplitude. Some indication of this is shown in
Fig.~\ref{fig:rms}, which includes a NLS1 galaxy PG 1404+226, which
clearly lies higher than the overall distribution. Thus it may be
possible to attribute the bulk of the scatter in Fig.~\ref{fig:rms} to
this other (but as yet unknown) parameter.

\subsubsection{Future prospects}

X-ray variability data have, until now, been collected in a rather
haphazard way, sometimes merely as a by-product of spectral studies.
This is unfortunate, as it makes it much harder to perform detailed,
systematic analysis. Also, the above correlations have only been
established conclusively for the very brightest sources, which are
predominantly low redshift, and on short time scales. Ideally, one
would like to obtain data over as wide a range of time scales as
possible, for as wide a range of luminosities as possible.

Medium-long time scale variability has been particularly neglected by
X-ray studies, mainly as it appeared initially that the most rapid
variations were the most interesting. This has not necessarily been
borne out by observations, and several groups are now using the
particular capabilities of \xte\ to obtain variability information on
time scales of months-to-years. Unfortunately, these experiments are
rather time-consuming and difficult to schedule. First results,
however, are promising. For example the PDS of NGC 3516, sampled from
hours-to-months, shows evidence for a cutoff on long time scales
(Edelson \& Nandra 1998) which had been suspected in other other
sources, but never measured explicitly (McHardy 1989; Papadakis \&
McHardy 1995). It is probable that this cutoff frequency is also
related to the luminosity. Similar data on a large sample of objects
is required to demonstrate this.

\begin{figure}[t]
\plotone{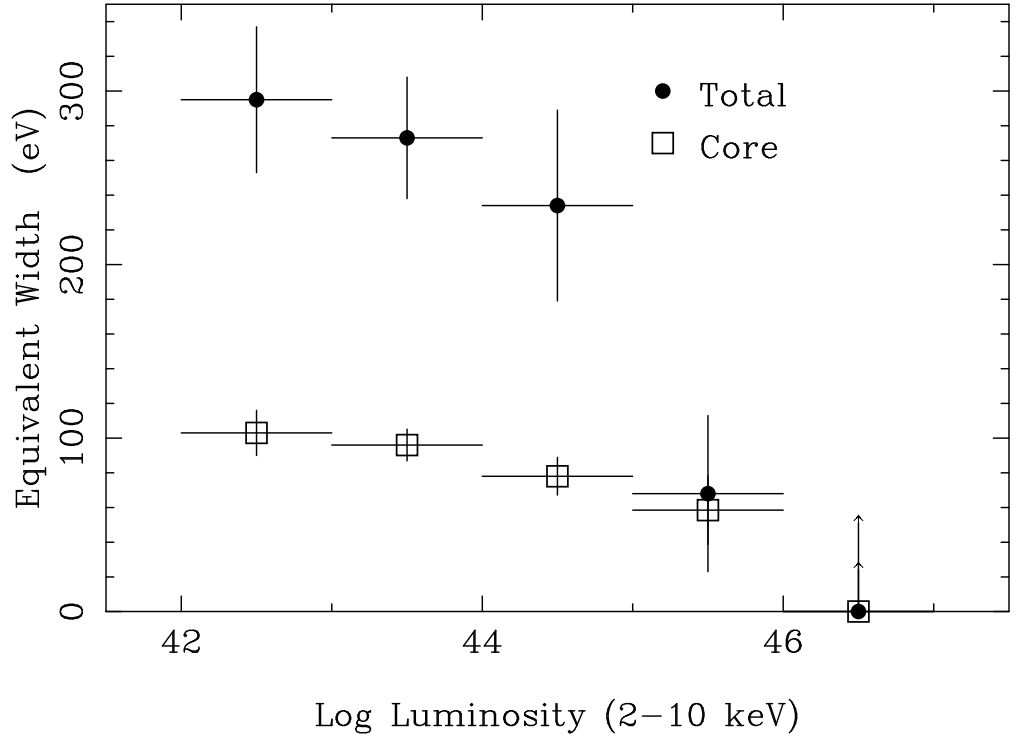}
\caption{
Equivalent width of the \fek\ emission lines versus \lx\ for a
sample of AGN observed by \asca.  The data are binned every decade of
luminosity.  Both the total EW and that of the narrow ``core'' reduce
with \lx.  Above $10^{45}$~erg s$^{-1}$ the core and total EWs
are consistent, showing the lack of any broad, red wing in the data.
\label{fig:lum_ew}}
\end{figure}

A further problem with rapid variability studies is that, mainly
due to the correlation described, high luminosity sources show
very small amplitudes of variability, which are consequently
extremely difficult to detect. The solution to this lies not in
more sensitive instrumentation, but rather in switching
attention to longer time scales, where all sources show higher
amplitudes of variability.

An exciting prospect for the future of long-time scale monitoring of
AGN is the employment of lobster-eye optics for X-ray imaging (Angel
1979; for a recent review see Peele \etal\ 1998). The idea behind this
technology is to provide an extremely large field-of-view (up to
all-sky), with arc-minute imaging. The more ambitious lobster-eye
mission concepts would be capable of monitoring 1000s of AGN
simultaneously on time scales of days to years, covering a wide range
of redshifts (Priedhorsky, Peele \& Nugent 1996). With very uniform
data over the right time scales for a large sample of AGN, such 
experiments offer the opportunity of revolutionizing our knowledge of
X-ray variability.

\subsection{\fek\ strength and profile}

\begin{figure}
\plotone{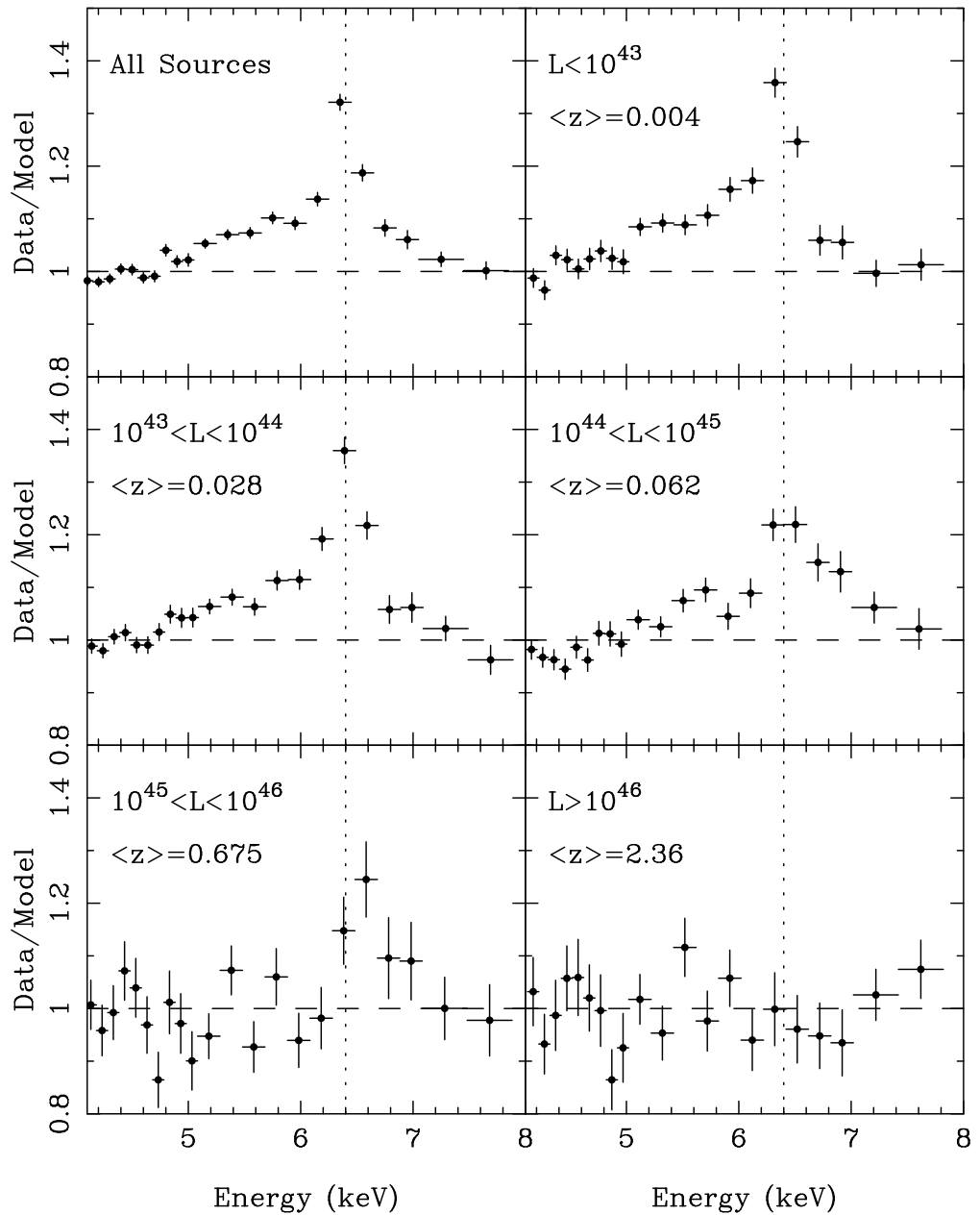}
\caption{Mean line profiles (expressed as the ratio of the data to a local
power-law model) of AGN observed by \asca, split into several
luminosity bins. The line strength and profile change above
\lx $\sim 10^{44}$~erg s$^{-1}$, with a reduction in both the core
and ``red wing''. Above $10^{46}$~erg s$^{-1}$ there is no evidence
for line emission at all.  
\label{fig:profs_lumin}}
\end{figure}

Iwasawa \& Taniguchi (1993), based on data from {\it Ginga}, suggested
that there may be an X-ray ``Baldwin'' effect, whereby the equivalent
width (EW) of the \fek\ line reduces with luminosity. This
result has recently been confirmed with much higher significance with
\asca\ (Nandra \etal\ 1997c; Fig~\ref{fig:lum_ew}). 
As mentioned in the introduction, the \fek\ lines in low-luminosity
AGN are complex, with a core peaking close to 6.4~keV and a strong and
very broad ``red wing''. The top left panel of
Fig.~\ref{fig:profs_lumin} shows the summed line profile for the
sample of AGN presented by Nandra \etal\ (1997c), illustrating these
features. This figure also illustrates the fact that the profile, as
well as then strength of the \fek\ line changes with luminosity.  The
bin with $10^{44}<$\lx$<10^{45}$ shows a weakening of the line
emission, especially at the core, and the the blue flux is relatively
stronger. At $10^{45}<$\lx$<10^{46}$ the red wing seems to have
disappeared, and the core now occurs at an energy higher then 6.4
keV. The total flux is even weaker now. Above $10^{46}$~erg s$^{-1}$
there is no evidence for any line emission at all.

\subsubsection{The origin of the correlation}

The \asca\ sample is not well-selected, and at least two important
parameters are also correlated with the X-ray luminosity: the redshift
and the radio-loudness. It is possible that, in fact, it is these
which are related to the equivalent width of the \fek\ line, rather
than the luminosity. Conclusive proof requires larger and
better-selected samples with data at higher signal-to-noise
ratios. For the time being, however, it seems most likely that the
luminosity is the primary parameter.  Although radio-loudness could be
an important factor in determining \fek\ strength, for example if the
X-rays are produced in a relativistic jet boosted away from the
accretion disk, it is unlikely to be the {\it sole} source of the
observed correlation, as it persists when only radio-quiet AGN are
considered. Similarly, although the very highest-luminosity bin with
no line emission contains only objects at very high redshift, there
are clear changes in profile in the lower-luminosity bins, where the
range of redshifts is very small.

If luminosity is indeed the important factor, an attractive
explanation of the observed EW and profile variations is that the
ionization state of the accretion disk increases with X-ray
luminosity. This would be expected with a fixed size and density.  As
the ionizing radiation field becomes more intense, we would expect the
iron atoms in the inner disk to become fully stripped, causing the
``red wing'' to disppear. Further out, high ionization species of iron
would cause a peak at higher energies than the 6.4~keV expected for
neutral iron. At extreme luminosities, iron in the X-ray illuminated
part of the disk could become completely stripped, with no line
emission being observed at all. If true, this suggests that
higher-luminosity AGN have higher accretion rates.

\subsubsection{Sources of scatter}

The \asca\ data are of insufficient quality to determine whether or
not there is significant scatter in the EW vs. \lx\ relation. From a
theoretical perspective, one would expect scatter.  For example, the
equivalent width of the \fek\ line is a strong function of disk
inclination (e.g., George \& Fabian 1991). The X-ray source and disk
geometry - in particular the solid angle of the disk subtended at the
X-ray source, is also related to the strength of the line. Further
speculation, however, requires improved data.

\subsubsection{Future prospects}

The emission lines in the low luminosity sources are reasonably well
measured by \asca, and although higher-resolution future missions such
as ASTRO-E will improve these measurements dramatically, what is
really required is to obtain reliable measurements and upper limits
for the high luminosity sources. This requires an instrument with
moderate resolution and high throughput, such as the XMM/EPIC.
Observations of large quasar samples - preferably with well-defined
selection criteria - will enable disentanglement of the various
related parameters and show whether or not luminosity is indeed the
driving force for the EW variations. One can also then search for
scatter in the X-ray Baldwin relation and examine its origins.

\section{The suitability of X-ray correlations for cosmology}

A summary of the various correlations of X-ray properties with
luminosity is given in Table~\ref{tab:summary}. As this volume
addresses the suitability of using QSO properties for cosmology,
the usefulness of these X-ray relations is discussed here.

\begin{table}
\begin{tabular}{llll}
\hline
Quantity & Satellite & Dependence & Reference \\
\hline
\hline
\aox\ & Einstein & $L^{-0.1}$ & Avni \& Tananbaum (1986) \\
\aox\ & {\it ROSAT} & $L^{}$ & Green \etal\ (1995) \\
$A_{\rm PDS}$ & {\it EXOSAT} & $L^{-0.5}$ & Lawrence \& Papadakis (1993) \\
$\sigma^{2}_{\rm RMS}$ & {\it ASCA} & $L^{-0.7}$ & Nandra \etal\ (1997a) \\
EW (Fe K$\alpha$) & Ginga & $L^{-0.2}$ & Iwasawa \& Taniguchi (1993) \\
EW (Fe K$\alpha$) & {\it ASCA} & $L^{-0.2}$ & Nandra \etal\ (1997c) \\ \hline
\end{tabular}
\caption{Summary of correlations \label{tab:summary}}
\end{table}

As can be seen from the Table, and a comparison with the traditional
Baldwin relations, it can be seen that the X-ray dependencies with
luminosity can be very strong. In particular both the variability
amplitude and \fek\ EW are very sharp functions of luminosity.  This
makes the relations more useful, as the data do not have to be
measured to so high an accuracy. The variability amplitude is perhaps
most promising of the relationships to use to calibrate cosmological
distances. Unfortunately, however, the relation is so strong that it
has not yet been possible to measure any variability at all in the
highest-luminosity (and therefore highest redshift) sources. As
mentioned above, a serious lobster-eye experiment could change that.
Fig.~\ref{fig:rms} does show scatter, but it is relatively small
compared to the traditional Baldwin relation. Although the origin of
the scatter is not understood it seems clear that at least some of it
arises from ``eigenvector 1'' and an attempt could be made to correct
for this scatter, using for example optical line width or soft X-ray
slope.

\section{Conclusions}

The correlation of X-ray properties with luminosity can reveal much
about the physics of AGN, particularly in the central regions. Their
relevance to cosmology is, at this stage, less certain as high-quality
data are only just beginning to emerge. Of the observed correlations,
the most promising is that of variability amplitude. This quantity has
a very steep dependence with luminosity and it may be possible to
reduce the scatter in the relationship using optical or X-ray spectral
data. All that remains is to assemble the necessary data, which may
be feasible with future, all-sky monitoring experiments, such as
lobster-eye optics.

\acknowledgments

Much of the credit for the above work goes to my collaborators on the
\asca\ projects, Ian George, Jane Turner, Richard Mushotzky and Tahir
Yaqoob. I thank Rick Edelson and Iossif Papadakis for many discussions
regarding X-ray variability. The potential of lobster-eye optics for
AGN studies was highlighted through discussion with Andrew Peele.
This work has been supported financially by the National Research
Council, Universities Space Research Association and NASA Grant
NAG5-7067.


\begin{references}
\reference Angel, J.R.P., 1979, ApJ, 233, 364
\reference Avni, Y., Tananbaum, H., 1982, ApJ, 262, L17
\reference Avni, Y., Tananbaum, H., 1986, ApJ, 305, 83
\reference Baldwin, J.A., 1977, ApJ, 214, 679
\reference Bao, G., Ambramowicz, M., 1996, ApJ, 465, 646
\reference Barr, P., Mushotzky, R.F., 1986, Nat, 320, 421
\reference Boller, Th., Brandt, W.N., Fink, H., 1996, A\& A 85, L11
\reference Boroson, T.A., Green, R.F., 1992, ApJS, 80, 109
\reference Edelson, R.A., Nandra, K., 1998, ApJ, submitted
\reference Fabian, A.C., Rees, M.J., Stella, L., White, N.E., 1989,
		MNRAS, 238, 729
\reference George, I.M., Fabian, A.C., 1991, MNRAS, 249, 352
\reference Green, A.R., McHardy, I.M., Lehto, H., 1993, MNRAS, 265, 644
\reference Green, P.J., \etal, 1995, ApJ, 450, 51
\reference Iwasawa, K., Taniguchi, Y., 1993, ApJ, 413, L15
\reference Kriss, G.A., Canizares, C.R., 1985, ApJ, 297, 177
\reference Laor, A., 1991, ApJ, 376, 90
\reference Laor, A., Fiore, F., Elvis, M., Wilkes, B.J., McDowell, J.C.,
		1997, ApJ, 477, 93
\reference Lawrence, A., Elvis, M., Wilkes, B.J., McHardy, I., Brandt, W.N.,
		1997, MNRAS, 285, 879
\reference Lawrence, A., Papadakis, I.E., 1993, ApJ, 41, L93
\reference McHardy, I.M., 1989, in ``Two Topics in X-ray astronomy'' Eds.,
		Hunt, H., Battrick, B., ESA SP-296 (ESA: Nordwijk), p. 1111  
\reference Malkan, M.A., Sargent, W.L., 1982, \apj, 254, 22
\reference Nandra, K., George, I.M., Mushotzky, R.F., Turner, T.J.,
		Yaqoob, T., 1997a, ApJ, 476, 70
\reference Nandra, K., George, I.M., Mushotzky, R.F., Turner, T.J.,
		Yaqoob, T., 1997b, ApJ, 477, 602
\reference Nandra, K., George, I.M., Mushotzky, R.F., Turner, T.J.,
		Yaqoob, T., 1997c, ApJ, 488, L91
\reference Nandra, K., Pounds, K.A., 1994, MNRAS, 268, 405
\reference Papadakis, I.E., McHardy, I.M., 1995, MNRAS, 273, 923
\reference Peele, A.G., \etal, 1998, Proc. SPIE 3334, in press
\reference Priedhorsky, W.C., Peele, A.G., Nugent, K.A., 1996, 
		 MNRAS 279, 733
\reference Tanaka, Y., \etal, 1995, Nat, 375, 659
\reference Turner, T.J., Pounds, K.A., 1989, MNRAS< 240, 833
\reference Wilkes, B.J., Tananbaum, H., Worral, D.M., Avni, Y., Oey, M.S.,
		Flanagan, J., 1994, ApJS, 92, 53
\reference Yuan, W., Siebert, J., Brinkmann, W., 1998, A\& A, 334, 498
\end{references}
\end{document}